\documentclass[journal=nalefd,manuscript=article]{achemso}
\setkeys{acs}{articletitle = true}

\usepackage[version=3]{mhchem} 
\usepackage{physics}

\author{Tom Dvir}
\affiliation{The Racah Institute of Physics, The Hebrew University of Jerusalem, Givat Ram, Jerusalem 91904
Israel}

\author{Marco Aprili}
\affiliation{Laboratoire de Physique des Solides (CNRS UMR 8502), Batiment 510, Universite Paris-Sud/Universite Paris-Saclay, 91405 Orsay, France}

\author{Charis H. L. Quay}
\affiliation{Laboratoire de Physique des Solides (CNRS UMR 8502), Batiment 510, Universite Paris-Sud/Universite Paris-Saclay, 91405 Orsay, France}

\author{Hadar Steinberg}
\affiliation{The Racah Institute of Physics, The Hebrew University of Jerusalem, Givat Ram, Jerusalem 91904
Israel}

\title{Zeeman tunability of Andreev bound states in van-der-Waals tunnel barriers} 


\begin{document}

\begin{abstract}
Quantum dots proximity-coupled to superconductors are attractive research platforms due to the intricate interplay between the single-electron nature of the dot and the many body nature of the superconducting state. These have been studied mostly using nanowires and carbon nanotubes, which allow a combination of tunability and proximity. Here we report a new type of quantum dot which allows proximity to a broad range of superconducting systems. The dots are realized as embedded defects within semiconducting tunnel barriers in van-der-Waals layers. By placing such layers on top of thin NbSe$_2$, we can probe the Andreev bound state spectra of such dots up to high in-plane magnetic fields without observing effects of a diminishing superconducting gap. As tunnel junctions defined on NbSe$_2$ have a hard gap, we can map the sub-gap spectra without background related to the rest of the junction. We find that the proximitized defect states invariably have a singlet ground state, manifest in the Zeeman splitting of the sub-gap excitation. We also find, in some cases, bound states which converge to zero energy and remain there. We discuss the role of the spin-orbit term, present both in the barrier and the superconductor, in the realization of such topologically trivial zero-energy states.
\end{abstract}

\section{Introduction}

In a quantum dot (QD) residing at close proximity to a superconductor, the excitation spectrum is governed by an interplay between induced superconductivity, charging energy, and chemical potential. Such coupling was initially studied by integrating dots into Josephson junctions (`S-QD-S' Devices) which may be studied in the strong or weak coupling regime~\cite{DeFranceschi:2010ce}. In the alternative `N-QD-S' geometry, sub-gap energies are probed directly. Here, the dot is weakly coupled to a normal electrode on one side, and strongly coupled to a superconductor on the other. Charge transfer through the dot and into the superconductor is carried through Andreev processes involving transitions between the ground and excited states~\cite{DeFranceschi:2010ce, Fazio:1998dl,Clerk:2000ge,Bauer:2007br,Governale:2008kp}. 
These transition energies appear as features in the tunneling spectra, below the superconducting gap $\Delta$. `N-QD-S' systems were realized by evaporating contacts on top of carbon nanotubes~\cite{Graber:2004cs,Pillet:2013hn}, self-assembled dots~\cite{Deacon:2010ka}, and semiconducting nanorwires (NWs)~\cite{Lee:2014gj,Jellinggaard:2016ig}. These systems allow for
gate-tunability of the dot chemical potential, generating a transition between two distinct ground states: An even parity, Cooper-pair-like singlet, and an odd parity, single-electron doublet~\cite{Meng:2009gm,Bauer:2007br, Deacon:2010jna, Lee:2014gj, Jellinggaard:2016ig}. Tuning the ground state into the doublet ground state is also possible by the application of in-plane magnetic field. In this case, the doublet state energy is Zeeman split, with the lower energy branch crossing the singlet energy at a finite applied field~\cite{Lee:2014gj}.

In recent years, a major research drive is aimed at probing the spectra of Majorana excitations, predicted to appear and observed as a zero-bias spectral feature in NWs proximity coupled to superconductors~\cite{Oreg:2010gk,Lutchyn:2010hp,Mourik2012a, Deng,  Das2012}.
Following these works, it became apparent that dots copuled to superconductors can also exhibit a peak similar to the expected Majorana signal at near zero energies. This happens when the dot is characterized by a strong spin-orbit coupling (SOC) term~\cite{Reeg:2018en,Liu:2017fb,Avila:2018vw}. 
More generally, understanding how the ABS spectrum develops at the presence of a SO term is important for distinguishing between trivial and topological states. 

A trivial system can exhibit a zero-energy spectral feature when the lower spin branch of the Zeeman split doublet state becomes degenerate with the singlet. Observing this crossover is an experimental challenge: It requires the energy scale $g\mu_BH$ (where $g$ is the Land\'e $g$ factor, $\mu_B$ is the Bohr magneton and $H$ is the magnitude of the field) to become significant while the superconductor retains a finite gap. To observe such splitting, some studies employed materials with a high $g$ factor~\cite{Lee:2014gj}, although results might be obscured by the diminishing of $\Delta$ with $H$. Here we use an alternative - to couple a quantum dot to an ultrathin superconductor - such as NbSe$_2$. NbSe$_2$ is a van-der-Waals superconductor, which remains superconducting at the ultrathin limit. Coupling QDs to NbSe$_2$ has two advantages: First, the superconducting gap of ultrathin NbSe$_2$ experiences very little change up to fields of a few Tesla in the plane~\cite{Dvir:2018cra}. Second, NbSe$_2$ and related transition metal dichalcogenide (TMD) materials are characterized by strong Ising SO coupling. It is of interest to consider the role of such SO terms on proximitized dot spectra.

In this work we fabricate and measure tunnel devices consisting of TMD barriers placed on top of NbSe$_2$, as reported in our earlier works~\cite{Dvir:2018fj, Dvir:2018cra}. The tunneling spectra exhibit Andreev bound states formed in naturally occurring quantum dots in the barrier. The spectra undergo clear Zeeman splitting at the presence of in-plane magnetic field, and are tracked up to 9 T. The majority of dots studied exhibit continuous spectral evolution, with a singlet to doublet crossover at some finite field. In some cases, however, we find a field-dependent transition to zero energy peaks. We suggest these are of trivial topology, and discuss their possible origin.

\section{Results}

\subsection{Observation of subgap states}
\begin{figure}
\includegraphics[width = 0.9\textwidth]{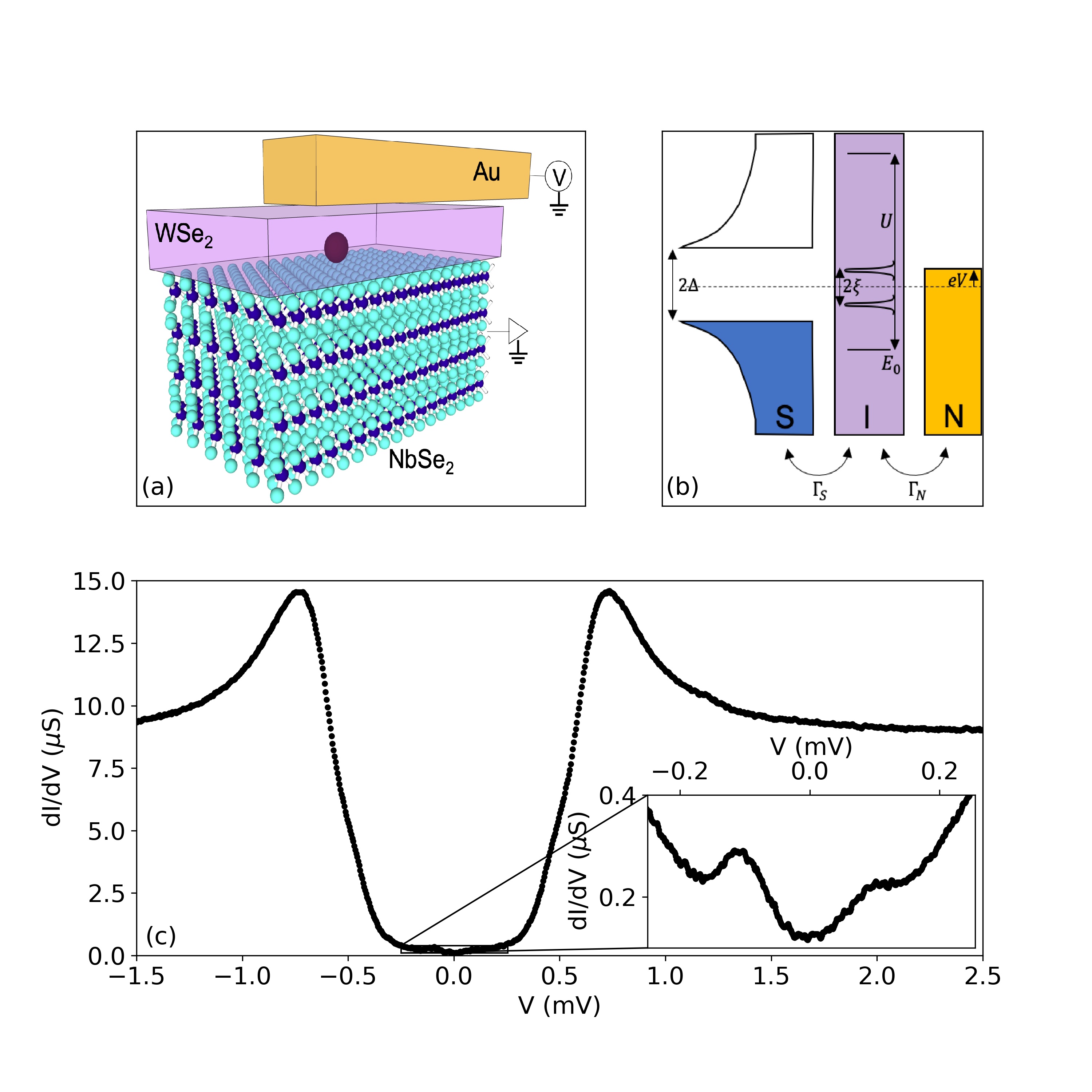}
\caption{\textbf{Sub gap states in van der Waals tunnel junctions: }\textbf{a.} Device schematics: a layer of NbSe$_2$ is connected to the ground and is covered by a thin barrier in which a quantum dot is formed. Ti/Au electrode is evaporated above and connected to a voltage source.  
\textbf{b.} Schematic model for tunneling between a superconductor to a normal metal through a quantum dot. Details in main text. \textbf{c.} Differential conductance of Device 1 at base temperature at zero magnetic field. Inset: magnification of the sub-gap region showing the two Andreev peaks formed symmetrically around zero.} 
\label{subgapstates}
\end{figure}

Figure \ref{subgapstates}a shows a sketch of the devices reported in this work. Normal - insulator - superconductor (NIS) tunnel junctions were fabricated using the dry transfer technique, by placing a few layer semiconductor TMD (WSe$_2$) on top of a flake of 2H-NbSe$_2$ (NbSe$_2$) of thickness ranging between 2 and 50 nm. Normal counter electrodes were fabricated using standard e-beam lithography methods as reported earlier \cite{Dvir:2018fj}. Typical junction dimensions are in the order of 1-2 $\mu$m$^2$ and barriers are 2-3 nm thick. Measurements are conducted using standard lock-in technique, where a bias voltage $V$ is applied to the Au counter electrode and the current $I$ and differential conductance $dI/dV$ are measured through a current pre-amp in ohmic contact with the NbSe$_2$ bulk. Measurements were conducted at base temperature below 70 mK, with AC excitation in the range of 20-50 $\mu$eV. 

The problem of a quantum dot that is coupled to a superconductor and to a normal metal is usually formulated in terms of the Anderson impurity model. This model accounts for tunneling between the dot and respective electrodes, and for the on-site electrostatic repulsion on the dot. The full solution of the model requires sophisticated computational tools and is beyond the scope of this paper. However we gain sufficient intuition by considering the case of a `superconducting impurity'\cite{Bauer:2007br,Baranski:2013gc}: when the coupling of the the superconducting electrode $\Gamma_S$ is much larger than the coupling to the normal electrode $\Gamma_N$ and the intrinsic superconducting gap $\Delta$ is the largest parameter in the system, an effective on-site interaction forms on the dot whose magnitude equals $\Delta_d = \Gamma_S/2$. The effective Hamiltonian then reads: 
\begin{equation}
    H_{QD} = \sum_\sigma E_{0} d^\dagger_\sigma d_\sigma - \frac{\Gamma_s}{2} \left(d^\dagger_\downarrow d^\dagger_\uparrow + h.c. \right) + U n_{d\downarrow} n_{d\uparrow}
\end{equation}
were $E_0$ is the energy level of the dot, $d^\dagger_\sigma,d_\sigma$ are the creation and annihilation operators for spin $\sigma$, $n_\sigma = d^\dagger_\sigma d_\sigma$ is the number operator for spin $\sigma$, and $U$ is the electrostatic repulsion energy of the dot, as depicted in the scheme shown in figure \ref{subgapstates}b.  Diagonalization of this Hamiltonian is straight-forward: There are two degenerate doublet eigenstates, $\ket{\uparrow}$, $\ket{\downarrow}$ with the energy $E_0$ and two `singlet' eigenstates which consist of the superposition of zero occupancy state, $\ket{0}$, and the double occupancy state, $\ket{\uparrow\downarrow}$ : 
\begin{equation}
    \ket{\Psi_-} = u_d \ket{0} - v_d \ket{\uparrow\downarrow}
\end{equation}
\begin{equation}
    \ket{\Psi_+} = v_d \ket{0} + u_d \ket{\uparrow\downarrow}
\end{equation}
\begin{equation}
    E_\pm = \left(E_0 + \frac{U}{2} \right) \pm \sqrt{\left(E_0 + \frac{U}{2} \right)^2 + \left(\frac{\Gamma_s}{2} \right)^2}
\end{equation}
The ground state of the system can be either the doublet or $\Psi_-$, depending on the interplay between $U$, $E_0$ and $\Gamma_s$. Tunneling experiments probe the energies corresponding to transitions where the number of electrons in the system is changed by one. In superconducting dots this process requires a transition between singlet and doublet states, with the energies: $\pm\xi = \frac{U}{2}  - \sqrt{\left(E_0 + \frac{U}{2} \right)^2 + \left(\frac{\Gamma_s}{2} \right)^2}$. While the simplified picture presented here is correct only in the limit of large superconducting gap, and doesn't take into account Kondo correlations, we believe that it qualitatively accounts for the observed data.

Figure \ref{subgapstates}c shows the differential conductance as measured with Device 1. The observed density of states shows a clear superconducting gap with quasi-particle peaks at energies of $\approx 800 \mu$V as discussed elsewhere \cite{Dvir:2018cra}. The sub-gap spectrum shows two peaks with energies of $\approx 100 \mu$V above a parabolic background. Such sub-gap peaks, observed in many of the tunnel junctions fabricated, are the subject of this report, and represent the singlet to doublet transition energy $\xi$. 

\begin{figure}
\includegraphics[width = 0.9\textwidth]{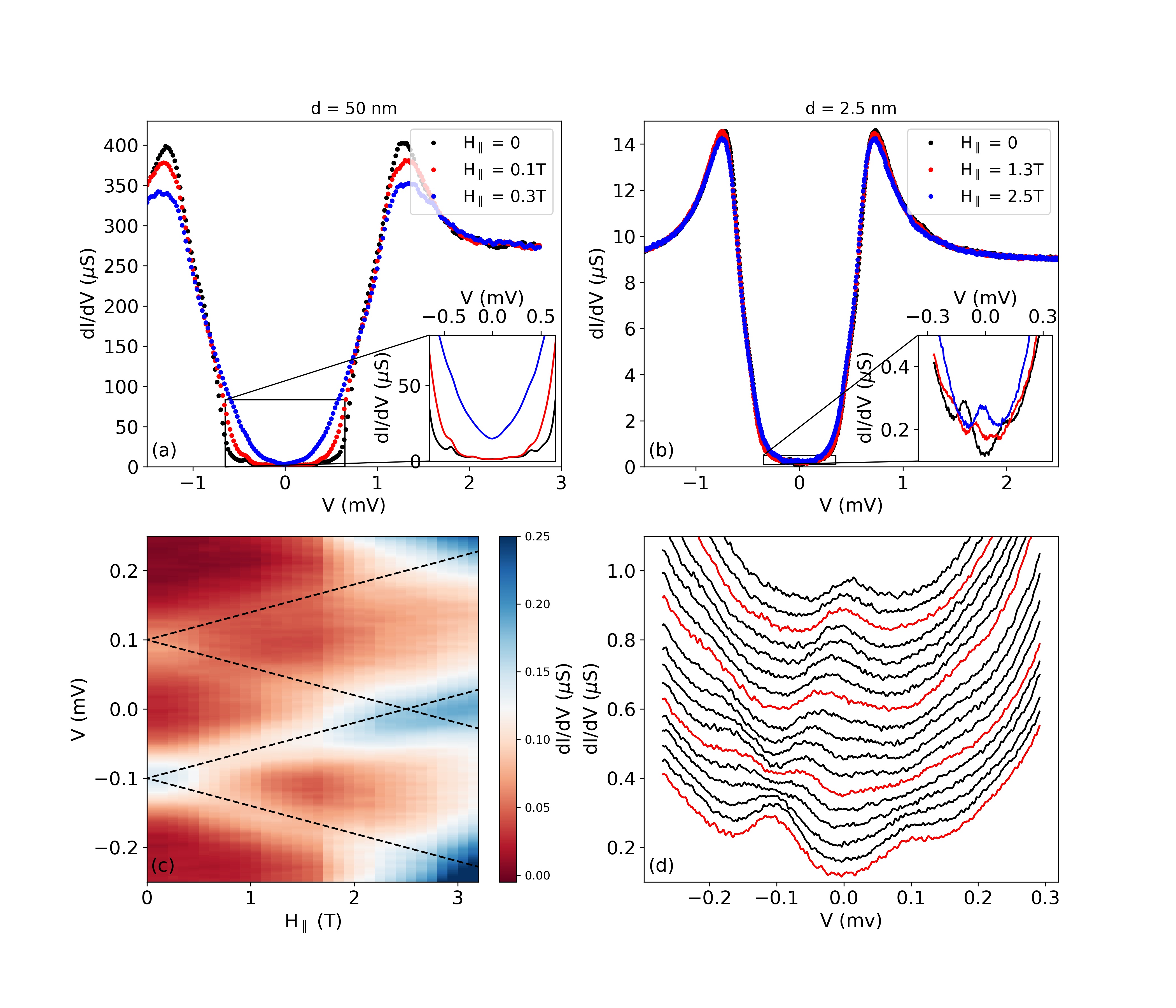}
\caption{\textbf{Sub gap states in van der Waals tunnel junctions: }\textbf{a.,b.} Differential conductance of Device 2 (50 nm thick) and Device 1 (2.5 nm thick) at increasing in-plane magnetic field. Insets: zoom in on the sub-gap spectrum.  \textbf{c.} Color map of the sub-gap conductance of Device 1 with increasing in-plane magnetic field. Dotted black lines trace the evolution of the sub-gap peaks with a slope of 40 $\mu$eV/T, equivalent to a $g$ factor of 1.3. \textbf{d.} Shifted differential conductance curves from the sub-gap region of Device 2, the curves are shown at intervals of 0.2 T. Red curves are shown at intervals of 1 T.} 
\label{fieldstates}
\end{figure}

\subsection{Magnetic field dependence}
An important knob for the control and study of dot-bound states is magnetic field. Application of field lifts the degeneracy between the two doublet states, with a Zeeman energy $E_z = \pm g \mu_B H$. Lifting of the doublet degeneracy allows the distinction between a singlet and a doublet ground state. When the ground-state is a doublet, application of field shifts its energy downwards, increasing the doublet-singlet energy difference and thus merely increasing the observed excitation energy. If, however, the ground-state is the singlet state, the excitation energy splits, eventually leading to a cross-over when the Zeeman energy equals $\xi$. Figure \ref{fieldstates}a shows the spectrum of Device 2 at in-plane magnetic fields between 0 and 0.3 T. While this field is insufficient for the observation of Zeeman effect, it is enough to allow for penetration of vortices whose spectroscopic signature overwhelm any other sub-gap features \cite{Dvir:2018fj}.  

To overcome this issue, we recall that few layer NbSe$_2$
has strong spin-orbit coupling, thus application of in-plane magnetic fields has negligible effect on the spectrum of such superconductors \cite{Dvir:2018cra}. Figure \ref{fieldstates}b shows the spectrum of Device 1, consisting of a tunnel junction into a 4 layer NbSe$_2$ flake, at in-plane magnetic fields ranging between 0 and 2.5 T. While the major features of the superconducting gap seem almost untouched by the field, the sub-gap spectrum changes significantly, as seen in Figure \ref{fieldstates}c which follows the evolution of the sub-gap spectrum of Device 1 with in-plane magnetic field. It is clear that the two symmetric sub-gap peaks split. The black traces fit the peak position according to a Zeeman energy with a $g$ factor of 1.3 . This observation, of splitting of the sub-gap peaks, repeats itself in all of the devices measured with $g$ factors in the range of 1.3 to 2 (see supplementary information for sub-gap spectra of all of the devices measured). 

This observation lends support to the claim that the zero field ground state of the proximitized dot is the singlet state. This was previously observed in quantum dots formed at the edges of nano-wires and carbon nano-tubes, with careful control over $E_0$ using dielectric gate. Compared with these systems, the proximitized dots formed in the vdW barriers tend to have a singlet ground-state, which points to small charging energy or to $E_0$ in the close vicinity of the Fermi energy. Furthermore, the observed magnitude of the $g$ factor points to atomic defect, rather than a large quantum dot with an internal band structure that re-normalizes $g$. For such an atomic defect, the broken inversion symmetry that leads to Ising spin-orbit coupling is of no importance. Thus, the spin orientation of the electrons on the dot are free to interact with the in-plane magnetic field. 

When the Zeeman energy equals the zero field $\xi$, a degeneracy between the singlet and the lower energy doublet state occurs, giving rise to a zero energy conductance peak. Further increase of the magnetic field beyond this crossover field, leads to a shift of the ground state to the lower doublet state, a crossover whose spectroscopic signature is the disappearance of the higher energy split peaks \cite{Lee:2014gj}. This is accompanied by crossing of the lower energy split peaks. Since the conduction in the junctions reported here consists of both tunneling through the quantum dot and tunneling directly between the normal metal and the superconductor, the higher energy peaks tend to merge with the above-gap conduction, thus hindering the observation of the former spectroscopic signal. The latter signal -- crossing of the split peaks at zero energy -- was evident in many of the measured devices (supplementary figure 2).   

\begin{figure}
\includegraphics[width = 0.9\textwidth]{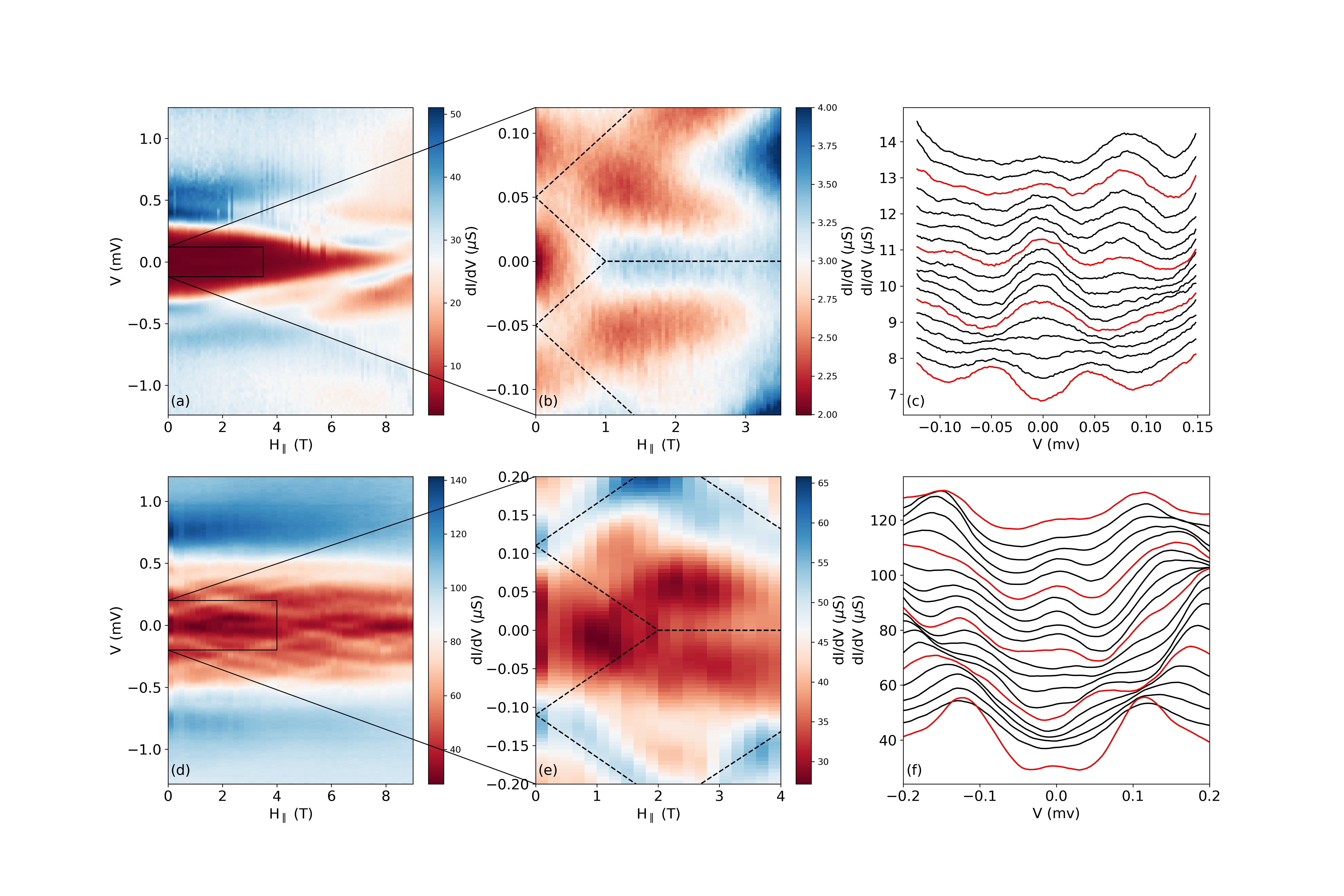}
\caption{\textbf{Formation of zero energy states: } \textbf{a.}  Color map of the sub-gap conductance of Device 3 with increasing in-plane magnetic field.  \textbf{b.} Close up on the sub-gap region showing the formation of a stable zero bias peak. \textbf{c.} Shifted differential conductance curves from the sub-gap region of Device 3, the curves are shown at intervals of 0.2 T. Red curves are shown at intervals of 1 T.  \textbf{d,e, f} Same for Device 4 } 
\label{ZBCP}
\end{figure}

\subsection{Zero bias conductance peaks (ZBCP)}
Figure 3 shows the differential conductance (panels a,c) and sub-gap conductance (panels b,d) of Devices 3 and 4 respectively.  In both devices a stable zero bias peak is formed at fields higher than the crossover field. This feature is stable for approximately 2.5 T, much higher than expected from spectral width of the sub-gap peaks. While in Device 3, the ZBCP is merged with the increasing background conductance beyond 3.5 T, in Device 4 the ZBCP splits at 4.5 T, only to reappear at 6.5 T. 

ABS pinning to zero energy, beyond some critical in-plane field, is a feature repeatedly seen in proximitized nanowires~\cite{Mourik2012a,Deng,Zhang:2018bx}. While such ZBCPs are often associated with Majorana fermions which appear due to non-trivial topology of the superconducting state, recent experimental \cite{Chen:2019us} and theoretical \cite{Reeg:2018en,Avila:2018vw,Liu:2017fb} studies point to trivial origins of zero energy pinning, calling for extra scrutiny of such results . A different source of zero bias peaks, originating in coupling between the quantum dot and the superconductor, has also been reported in proximitized nanowires \cite{Lee:2012ft,Jellinggaard:2016ig}.
In what follows, we discuss alternative interpretations for the emergence of topologically trivial ZBCPs, in the NIS van der Waals system.

In principle, the system discussed here possesses the required ingredients for the formation of topological superconductivity that breaks time reversal symmetry and Majorana bound states: superconductivity, strong spin orbit coupling, and magnetic field which is applied perpendicular to the direction of the SOC. Formation of topological superconductivity is unlikely, as topological systems require the proximitized region to be large in at least a single dimension, to enable the formation of edge states. Since the sub-gap features reported here are in all likelihood associated with atomic scale quantum dots, as evident in the observed low $g$ factor, we can rule out the topological origin of the stable ZBCP that is discussed in the context of nanowires \cite{Oreg:2010gk,Lutchyn:2010hp}. The combination of superconductivity, strong spin-orbit coupling and Zeeman field can form trivial nearly zero energy bound states in quantum dots, as recently shown in refs \cite{Reeg:2018en,Liu:2017fb}. There, a non superconducting quantum dot, in contact with a BCS superconductor, was theoretically shown to host such bound states when a magnetic field crosses a threshold, determined by the superconducting gap and the strength of the SOC. While the details of the discussed model and our van der Walls system are different, we believe that the phenomenon of pinning to zero bias is general. We study here a quantum dot embedded in a semi-conducting barrier that hosts a strong intrinsic Ising SOC in addition to Rashba SOC, in proximity to an ultra-thin Ising superconductor. This special type of proximitized quantum dot calls for further theoretical modelling.

A different possibility for the formation of zero bias peaks comes from the Kondo effect. It was shown that the ground state of a quantum dot coupled both to a superconductor and to a normal metal can be either the doublet or superconducting singlet as discussed, or a Kondo singlet that involves a superposition between the electrons in the dot and the electrons in the normal lead \cite{Jellinggaard:2016ig}. Application of magnetic field can induce a SC singlet to Kondo singlet transition as a result of reduction in the magnitude of the superconductor order parameter or filling of the SC gap \cite{Lee:2012ft}. Kondo peaks, however, are stable only in magnetic fields smaller than the Kondo temperature $T_K$, in which the Kondo resonance becomes apparent. Beyond such small fields, the Kondo degeneracy splits \cite{Deacon:2010ka}. Such splitting is not observed here.

Furthermore, the actual reduction in the magnitude of the superconducting gap can pin the excitation energy of the dot to zero by level repulsion\cite{Jellinggaard:2016ig}. Few layer NbSe$_2$ is very resilient to the application of in-plane field. In fields of the range reported in this work, the superconducting gap hardly changes \cite{Dvir:2018fj}, rendering both mechanisms - Kondo or repulsion from the gap  - implausible. 

\section{Conclusions}
In summary, we show that vdW tunnel junctions using TMD barriers may serve as a platform to study the proximity between quantum dots and superconductors. This platform is set in the extreme limit of  quantum dots whose dimensions are on the atomic scale, and also in the presence of a strong Ising spin orbit coupling. Our results suggest that the zero field ground state of such dots is analogous to the BCS singlet state, which can be tuned by the application of in-plane magnetic field. Finally, the formation of stable zero bias spectral features at finite magnetic fields calls for further theoretical investigation.

\section{Acknowledgements}
The authors are thankful for fruitful discussions with Y. Oreg and D. Loss. The author are also thankful for A. Zalic, A. Vakahi and S. Remennik for TEM and SEM imaging. This work was fundede by a Maimonïdes-Israel grant from the Israeli-French High Council for Scientific \& Technological Research, an ANR JCJC grant (SPINOES) from the French Agence Nationale de Recherche, a European Research Council Starting Grant (No. 637298, TUNNEL), and Israeli Science Foundation grant 1363/15. T.D. is grateful to the Azrieli Foundation for an Azrieli Fellowship. 

\section{Author contributions}
T.D. fabricated the devices. C.Q.H.L., T.D. and M.A. performed the measurements. All authors contributed to data analysis and the writing of the manuscript.

\section{Competing financial interests}
The authors declare no competing financial interests.

\bibliography{bibliography}

\providecommand{\latin}[1]{#1}
\makeatletter
\providecommand{\doi}
  {\begingroup\let\do\@makeother\dospecials
  \catcode`\{=1 \catcode`\}=2 \doi@aux}
\providecommand{\doi@aux}[1]{\endgroup\texttt{#1}}
\makeatother
\providecommand*\mcitethebibliography{\thebibliography}
\csname @ifundefined\endcsname{endmcitethebibliography}
  {\let\endmcitethebibliography\endthebibliography}{}
\begin{mcitethebibliography}{27}
\providecommand*\natexlab[1]{#1}
\providecommand*\mciteSetBstSublistMode[1]{}
\providecommand*\mciteSetBstMaxWidthForm[2]{}
\providecommand*\mciteBstWouldAddEndPuncttrue
  {\def\EndOfBibitem{\unskip.}}
\providecommand*\mciteBstWouldAddEndPunctfalse
  {\let\EndOfBibitem\relax}
\providecommand*\mciteSetBstMidEndSepPunct[3]{}
\providecommand*\mciteSetBstSublistLabelBeginEnd[3]{}
\providecommand*\EndOfBibitem{}
\mciteSetBstSublistMode{f}
\mciteSetBstMaxWidthForm{subitem}{(\alph{mcitesubitemcount})}
\mciteSetBstSublistLabelBeginEnd
  {\mcitemaxwidthsubitemform\space}
  {\relax}
  {\relax}

\bibitem[De~Franceschi \latin{et~al.}(2010)De~Franceschi, Kouwenhoven,
  Sch{\"o}nenberger, and Wernsdorfer]{DeFranceschi:2010ce}
De~Franceschi,~S.; Kouwenhoven,~L.; Sch{\"o}nenberger,~C.; Wernsdorfer,~W.
  {Hybrid superconductor{\textendash}quantum dot devices}. \emph{Nature
  Nanotechnology} \textbf{2010}, \emph{5}, 703--711\relax
\mciteBstWouldAddEndPuncttrue
\mciteSetBstMidEndSepPunct{\mcitedefaultmidpunct}
{\mcitedefaultendpunct}{\mcitedefaultseppunct}\relax
\EndOfBibitem
\bibitem[Fazio and Raimondi(1998)Fazio, and Raimondi]{Fazio:1998dl}
Fazio,~R.; Raimondi,~R. {Resonant Andreev Tunneling in Strongly Interacting
  Quantum Dots}. \emph{Physical Review Letters} \textbf{1998}, \emph{80},
  2913--2916\relax
\mciteBstWouldAddEndPuncttrue
\mciteSetBstMidEndSepPunct{\mcitedefaultmidpunct}
{\mcitedefaultendpunct}{\mcitedefaultseppunct}\relax
\EndOfBibitem
\bibitem[Clerk \latin{et~al.}(2000)Clerk, Ambegaokar, and
  Hershfield]{Clerk:2000ge}
Clerk,~A.~A.; Ambegaokar,~V.; Hershfield,~S. {Andreev scattering and the Kondo
  effect}. \emph{Physical Review B} \textbf{2000}, \emph{61}, 3555--3562\relax
\mciteBstWouldAddEndPuncttrue
\mciteSetBstMidEndSepPunct{\mcitedefaultmidpunct}
{\mcitedefaultendpunct}{\mcitedefaultseppunct}\relax
\EndOfBibitem
\bibitem[Bauer \latin{et~al.}(2007)Bauer, Oguri, and Hewson]{Bauer:2007br}
Bauer,~J.; Oguri,~A.; Hewson,~A.~C. {Spectral properties of locally correlated
  electrons in a Bardeen{\textendash}Cooper{\textendash}Schrieffer
  superconductor}. \emph{Journal of Physics: Condensed Matter} \textbf{2007},
  \emph{19}, 486211--20\relax
\mciteBstWouldAddEndPuncttrue
\mciteSetBstMidEndSepPunct{\mcitedefaultmidpunct}
{\mcitedefaultendpunct}{\mcitedefaultseppunct}\relax
\EndOfBibitem
\bibitem[Governale \latin{et~al.}(2008)Governale, Pala, and
  K{\"o}nig]{Governale:2008kp}
Governale,~M.; Pala,~M.~G.; K{\"o}nig,~J. {Real-time diagrammatic approach to
  transport through interacting quantum dots with normal and superconducting
  leads}. \emph{Physical Review B} \textbf{2008}, \emph{77}, 659--14\relax
\mciteBstWouldAddEndPuncttrue
\mciteSetBstMidEndSepPunct{\mcitedefaultmidpunct}
{\mcitedefaultendpunct}{\mcitedefaultseppunct}\relax
\EndOfBibitem
\bibitem[Gr{\"a}ber \latin{et~al.}(2004)Gr{\"a}ber, Nussbaumer, Belzig, and
  Sch{\"o}nenberger]{Graber:2004cs}
Gr{\"a}ber,~M.~R.; Nussbaumer,~T.; Belzig,~W.; Sch{\"o}nenberger,~C. {Quantum
  dot coupled to a normal and a superconducting lead}. \emph{Nanotechnology}
  \textbf{2004}, \emph{15}, S479--S482\relax
\mciteBstWouldAddEndPuncttrue
\mciteSetBstMidEndSepPunct{\mcitedefaultmidpunct}
{\mcitedefaultendpunct}{\mcitedefaultseppunct}\relax
\EndOfBibitem
\bibitem[Pillet \latin{et~al.}(2013)Pillet, Joyez, {\v Z}itko, and
  Goffman]{Pillet:2013hn}
Pillet,~J.-D.; Joyez,~P.; {\v Z}itko,~R.; Goffman,~M.~F. {Tunneling
  spectroscopy of a single quantum dot coupled to a superconductor: From Kondo
  ridge to Andreev bound states}. \emph{Physical Review B} \textbf{2013},
  \emph{88}, 045101--6\relax
\mciteBstWouldAddEndPuncttrue
\mciteSetBstMidEndSepPunct{\mcitedefaultmidpunct}
{\mcitedefaultendpunct}{\mcitedefaultseppunct}\relax
\EndOfBibitem
\bibitem[Deacon \latin{et~al.}(2010)Deacon, Tanaka, Oiwa, Sakano, Yoshida,
  Shibata, Hirakawa, and Tarucha]{Deacon:2010ka}
Deacon,~R.~S.; Tanaka,~Y.; Oiwa,~A.; Sakano,~R.; Yoshida,~K.; Shibata,~K.;
  Hirakawa,~K.; Tarucha,~S. {Kondo-enhanced Andreev transport in single
  self-assembled InAs quantum dots contacted with normal and superconducting
  leads}. \emph{Physical Review B} \textbf{2010}, \emph{81}, 659--4\relax
\mciteBstWouldAddEndPuncttrue
\mciteSetBstMidEndSepPunct{\mcitedefaultmidpunct}
{\mcitedefaultendpunct}{\mcitedefaultseppunct}\relax
\EndOfBibitem
\bibitem[Lee \latin{et~al.}(2014)Lee, Jiang, Houzet, Aguado, Lieber, and
  De~Franceschi]{Lee:2014gj}
Lee,~E. J.~H.; Jiang,~X.; Houzet,~M.; Aguado,~R.; Lieber,~C.~M.;
  De~Franceschi,~S. {Spin-resolved Andreev levels and parity crossings in
  hybrid superconductor{\textendash}semiconductor nanostructures}. \emph{Nature
  Nanotechnology} \textbf{2014}, \emph{9}, 79--84\relax
\mciteBstWouldAddEndPuncttrue
\mciteSetBstMidEndSepPunct{\mcitedefaultmidpunct}
{\mcitedefaultendpunct}{\mcitedefaultseppunct}\relax
\EndOfBibitem
\bibitem[Jellinggaard \latin{et~al.}(2016)Jellinggaard, Grove-Rasmussen,
  Madsen, and Nyg{\aa}rd]{Jellinggaard:2016ig}
Jellinggaard,~A.; Grove-Rasmussen,~K.; Madsen,~M.~H.; Nyg{\aa}rd,~J. {Tuning
  Yu-Shiba-Rusinov states in a quantum dot}. \emph{Physical Review B}
  \textbf{2016}, \emph{94}, 85--8\relax
\mciteBstWouldAddEndPuncttrue
\mciteSetBstMidEndSepPunct{\mcitedefaultmidpunct}
{\mcitedefaultendpunct}{\mcitedefaultseppunct}\relax
\EndOfBibitem
\bibitem[Meng \latin{et~al.}(2009)Meng, Florens, and Simon]{Meng:2009gm}
Meng,~T.; Florens,~S.; Simon,~P. {Self-consistent description of Andreev bound
  states in Josephson quantum dot devices}. \emph{Physical Review B}
  \textbf{2009}, \emph{79}, 659--10\relax
\mciteBstWouldAddEndPuncttrue
\mciteSetBstMidEndSepPunct{\mcitedefaultmidpunct}
{\mcitedefaultendpunct}{\mcitedefaultseppunct}\relax
\EndOfBibitem
\bibitem[Deacon \latin{et~al.}(2010)Deacon, Tanaka, Oiwa, Sakano, Yoshida,
  Shibata, Hirakawa, and Tarucha]{Deacon:2010jna}
Deacon,~R.~S.; Tanaka,~Y.; Oiwa,~A.; Sakano,~R.; Yoshida,~K.; Shibata,~K.;
  Hirakawa,~K.; Tarucha,~S. {Tunneling Spectroscopy of Andreev Energy Levels in
  a Quantum Dot Coupled to a Superconductor}. \emph{Physical review letters}
  \textbf{2010}, \emph{104}, 076805--4\relax
\mciteBstWouldAddEndPuncttrue
\mciteSetBstMidEndSepPunct{\mcitedefaultmidpunct}
{\mcitedefaultendpunct}{\mcitedefaultseppunct}\relax
\EndOfBibitem
\bibitem[Oreg \latin{et~al.}(2010)Oreg, Refael, and von Oppen]{Oreg:2010gk}
Oreg,~Y.; Refael,~G.; von Oppen,~F. {Helical Liquids and Majorana Bound States
  in Quantum Wires}. \emph{Physical Review Letters} \textbf{2010}, \emph{105},
  177002\relax
\mciteBstWouldAddEndPuncttrue
\mciteSetBstMidEndSepPunct{\mcitedefaultmidpunct}
{\mcitedefaultendpunct}{\mcitedefaultseppunct}\relax
\EndOfBibitem
\bibitem[Lutchyn \latin{et~al.}(2010)Lutchyn, Sau, and
  Das~Sarma]{Lutchyn:2010hp}
Lutchyn,~R.~M.; Sau,~J.~D.; Das~Sarma,~S. {Majorana Fermions and a Topological
  Phase Transition in Semiconductor-Superconductor Heterostructures}.
  \emph{Physical review letters} \textbf{2010}, \emph{105}, 077001--4\relax
\mciteBstWouldAddEndPuncttrue
\mciteSetBstMidEndSepPunct{\mcitedefaultmidpunct}
{\mcitedefaultendpunct}{\mcitedefaultseppunct}\relax
\EndOfBibitem
\bibitem[Mourik \latin{et~al.}(2012)Mourik, Zuo, Frolov, Plissard, Bakkers, and
  Kouwenhoven]{Mourik2012a}
Mourik,~V.; Zuo,~K.; Frolov,~S.~M.; Plissard,~S.~R.; Bakkers,~E. P. a.~M.;
  Kouwenhoven,~L.~P. {Signatures of Majorana fermions in hybrid
  superconductor-semiconductor nanowire devices.} \emph{Science} \textbf{2012},
  \emph{336}, 1003--1007\relax
\mciteBstWouldAddEndPuncttrue
\mciteSetBstMidEndSepPunct{\mcitedefaultmidpunct}
{\mcitedefaultendpunct}{\mcitedefaultseppunct}\relax
\EndOfBibitem
\bibitem[Deng \latin{et~al.}(2016)Deng, Vaitiek{\.{e}}nas, Hansen, Danon,
  Leijnse, Flensberg, Nyg{\aa}rd, Krogstrup, and Marcus]{Deng}
Deng,~M.~T.; Vaitiek{\.{e}}nas,~S.; Hansen,~E.~B.; Danon,~J.; Leijnse,~M.;
  Flensberg,~K.; Nyg{\aa}rd,~J.; Krogstrup,~P.; Marcus,~C.~M. {Majorana bound
  state in a coupled quantum-dot hybrid-nanowire system.} \emph{Science}
  \textbf{2016}, \emph{354}, 1557--1562\relax
\mciteBstWouldAddEndPuncttrue
\mciteSetBstMidEndSepPunct{\mcitedefaultmidpunct}
{\mcitedefaultendpunct}{\mcitedefaultseppunct}\relax
\EndOfBibitem
\bibitem[Das \latin{et~al.}(2012)Das, Ronen, Most, Oreg, Heiblum, and
  Shtrikman]{Das2012}
Das,~A.; Ronen,~Y.; Most,~Y.; Oreg,~Y.; Heiblum,~M.; Shtrikman,~H. {Zero-bias
  peaks and splitting in an Al{\textendash}InAs nanowire topological
  superconductor as a signature of Majorana fermions}. \emph{Nature Physics}
  \textbf{2012}, \emph{8}, 887--895\relax
\mciteBstWouldAddEndPuncttrue
\mciteSetBstMidEndSepPunct{\mcitedefaultmidpunct}
{\mcitedefaultendpunct}{\mcitedefaultseppunct}\relax
\EndOfBibitem
\bibitem[Reeg \latin{et~al.}(2018)Reeg, Dmytruk, Chevallier, Loss, and
  Klinovaja]{Reeg:2018en}
Reeg,~C.; Dmytruk,~O.; Chevallier,~D.; Loss,~D.; Klinovaja,~J. {Zero-energy
  Andreev bound states from quantum dots in proximitized Rashba nanowires}.
  \emph{Physical Review B} \textbf{2018}, \emph{98}, 1--12\relax
\mciteBstWouldAddEndPuncttrue
\mciteSetBstMidEndSepPunct{\mcitedefaultmidpunct}
{\mcitedefaultendpunct}{\mcitedefaultseppunct}\relax
\EndOfBibitem
\bibitem[Liu \latin{et~al.}(2017)Liu, Sau, Stanescu, and Das~Sarma]{Liu:2017fb}
Liu,~C.-X.; Sau,~J.~D.; Stanescu,~T.~D.; Das~Sarma,~S. {Andreev bound states
  versus Majorana bound states in quantum dot-nanowire-superconductor hybrid
  structures: Trivial versus topological zero-bias conductance peaks}.
  \emph{Physical Review B} \textbf{2017}, \emph{96}\relax
\mciteBstWouldAddEndPuncttrue
\mciteSetBstMidEndSepPunct{\mcitedefaultmidpunct}
{\mcitedefaultendpunct}{\mcitedefaultseppunct}\relax
\EndOfBibitem
\bibitem[Avila \latin{et~al.}(2018)Avila, Pe{\~n}aranda, Prada, San-Jose, and
  Aguado]{Avila:2018vw}
Avila,~J.; Pe{\~n}aranda,~F.; Prada,~E.; San-Jose,~P.; Aguado,~R.
  {Non-Hermitian topology: a unifying framework for the Andreev versus Majorana
  states controversy}. \emph{arXiv:1807.04677} \textbf{2018}, \relax
\mciteBstWouldAddEndPunctfalse
\mciteSetBstMidEndSepPunct{\mcitedefaultmidpunct}
{}{\mcitedefaultseppunct}\relax
\EndOfBibitem
\bibitem[Dvir \latin{et~al.}(2018)Dvir, Aprili, Quay, and
  Steinberg]{Dvir:2018cra}
Dvir,~T.; Aprili,~M.; Quay,~C. H.~L.; Steinberg,~H. {Tunneling into the Vortex
  State of NbSe$_{2}$ with van der Waals Junctions}. \emph{Nano Letters}
  \textbf{2018}, \emph{18}, 7845--7850\relax
\mciteBstWouldAddEndPuncttrue
\mciteSetBstMidEndSepPunct{\mcitedefaultmidpunct}
{\mcitedefaultendpunct}{\mcitedefaultseppunct}\relax
\EndOfBibitem
\bibitem[Dvir \latin{et~al.}(2018)Dvir, Massee, Attias, Khodas, Aprili, Quay,
  and Steinberg]{Dvir:2018fj}
Dvir,~T.; Massee,~F.; Attias,~L.; Khodas,~M.; Aprili,~M.; Quay,~C. H.~L.;
  Steinberg,~H. {Spectroscopy of bulk and few-layer superconducting NbSe$_2$
  with van der Waals tunnel junctions}. \emph{Nature Communications}
  \textbf{2018}, \emph{9}, 598\relax
\mciteBstWouldAddEndPuncttrue
\mciteSetBstMidEndSepPunct{\mcitedefaultmidpunct}
{\mcitedefaultendpunct}{\mcitedefaultseppunct}\relax
\EndOfBibitem
\bibitem[Bara{\'{n}}ski and Doma{\'{n}}ski(2013)Bara{\'{n}}ski, and
  Doma{\'{n}}ski]{Baranski:2013gc}
Bara{\'{n}}ski,~J.; Doma{\'{n}}ski,~T. {In-gap states of a quantum dot coupled
  between a normal and a superconducting lead}. \emph{Journal of Physics:
  Condensed Matter} \textbf{2013}, \emph{25}, 435305--10\relax
\mciteBstWouldAddEndPuncttrue
\mciteSetBstMidEndSepPunct{\mcitedefaultmidpunct}
{\mcitedefaultendpunct}{\mcitedefaultseppunct}\relax
\EndOfBibitem
\bibitem[Zhang \latin{et~al.}(2018)Zhang, Liu, Gazibegovic, Xu, Logan, Wang,
  van Loo, Bommer, de~Moor, Car, Veld, van Veldhoven, Koelling, Verheijen,
  Pendharkar, Pennachio, Shojaei, Lee, Palmstr{\o}m, Bakkers, Das~Sarma, and
  Kouwenhoven]{Zhang:2018bx}
Zhang,~H. \latin{et~al.}  {Quantized Majorana conductance}. \emph{Nature}
  \textbf{2018}, \emph{556}, 74--79\relax
\mciteBstWouldAddEndPuncttrue
\mciteSetBstMidEndSepPunct{\mcitedefaultmidpunct}
{\mcitedefaultendpunct}{\mcitedefaultseppunct}\relax
\EndOfBibitem
\bibitem[Chen \latin{et~al.}(2019)Chen, Woods, Yu, Hocevar, Car, Plissard,
  Bakkers, Stanescu, and Frolov]{Chen:2019us}
Chen,~J.; Woods,~B.; Yu,~P.; Hocevar,~M.; Car,~D.; Plissard,~S.; Bakkers,~E.;
  Stanescu,~T.; Frolov,~S. {Ubiquitous non-Majorana Zero-Bias Conductance Peaks
  in Nanowire Devices}. \emph{arXiv:1902.02773} \textbf{2019}, \relax
\mciteBstWouldAddEndPunctfalse
\mciteSetBstMidEndSepPunct{\mcitedefaultmidpunct}
{}{\mcitedefaultseppunct}\relax
\EndOfBibitem
\bibitem[Lee \latin{et~al.}(2012)Lee, Jiang, Aguado, Katsaros, Lieber, and
  De~Franceschi]{Lee:2012ft}
Lee,~E. J.~H.; Jiang,~X.; Aguado,~R.; Katsaros,~G.; Lieber,~C.~M.;
  De~Franceschi,~S. {Zero-Bias Anomaly in a Nanowire Quantum Dot Coupled to
  Superconductors}. \emph{Physical review letters} \textbf{2012}, \emph{109},
  659--5\relax
\mciteBstWouldAddEndPuncttrue
\mciteSetBstMidEndSepPunct{\mcitedefaultmidpunct}
{\mcitedefaultendpunct}{\mcitedefaultseppunct}\relax
\EndOfBibitem
\end{mcitethebibliography}

\pagebreak
\begin{center}
\textbf{\large Supplemental Materials: Zeeman tunability of Andreev bound states in van-der-Waals tunnel barriers}
\end{center}
\setcounter{equation}{0}
\setcounter{figure}{0}
\setcounter{table}{0}
\setcounter{page}{1}
\makeatletter
\renewcommand{\theequation}{S\arabic{equation}}
\renewcommand\thesection{S\arabic{section}}
\renewcommand\thefigure{\textbf{S\arabic{figure}}}   
\renewcommand{\figurename}{\textbf{Supplementary Figure}}
\setcounter{secnumdepth}{1}

\begin{figure}
\includegraphics[width = 0.9\textwidth]{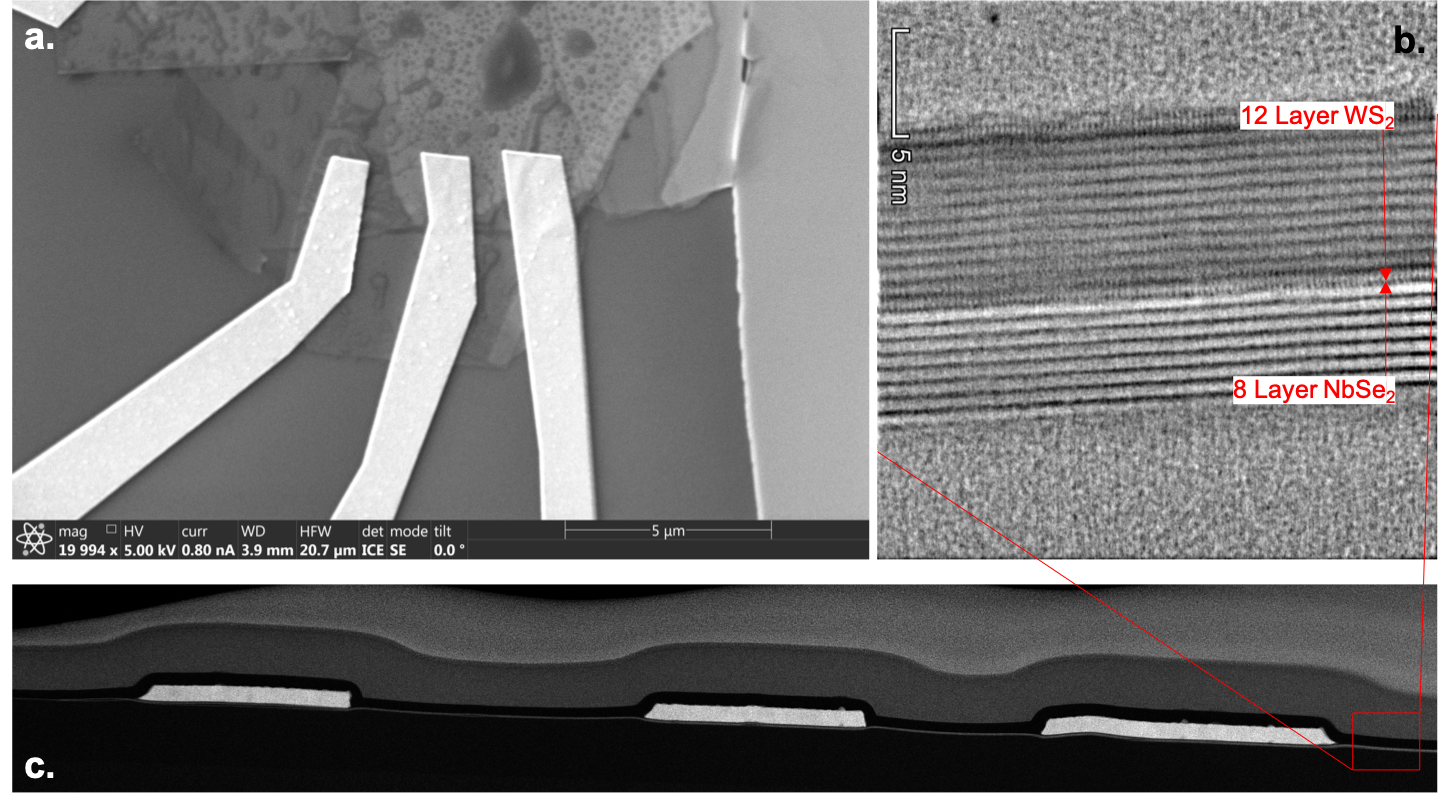}
\caption{\textbf{Typical device interface: a.} SEM imaging of Device 5. \textbf{b., c.} Cross section TEM along a line cut of Device 5.  } 
\label{sup: interface}
\end{figure}

\begin{figure}
\includegraphics[width = 0.9\textwidth]{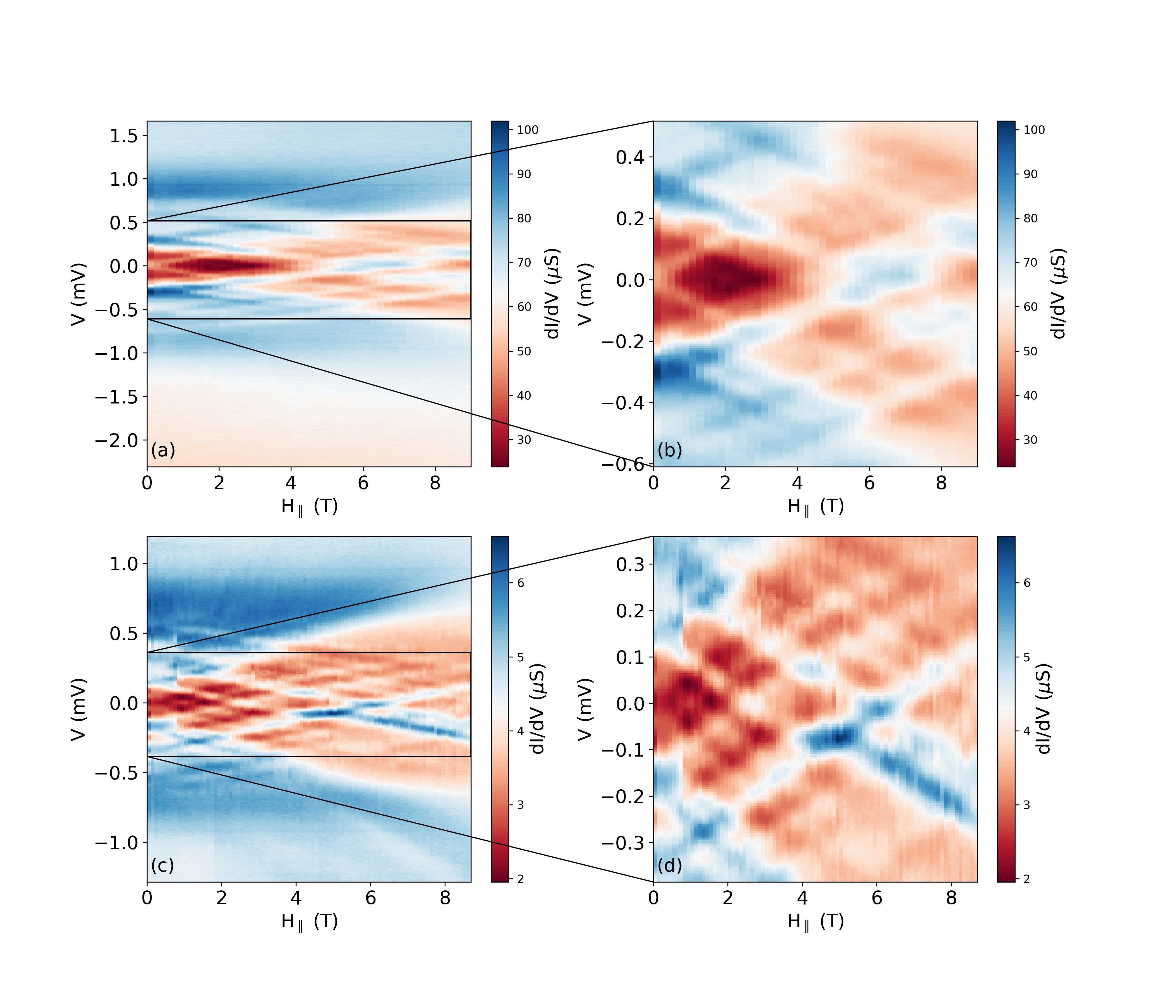}
\caption{\textbf{Additional devices: } \textbf{a}  Color map of the sub-gap conductance of Device 6 with increasing in-plane magnetic field.  \textbf{b.} Zoom in on the sub-gap region showing the linear dispersion of the energy of the sub-gap peaks with magnetic field. \textbf{c,d} Same for Device 7 } 
\label{sup: additional devices}
\end{figure}

\begin{figure}
\includegraphics[width = 0.9\textwidth]{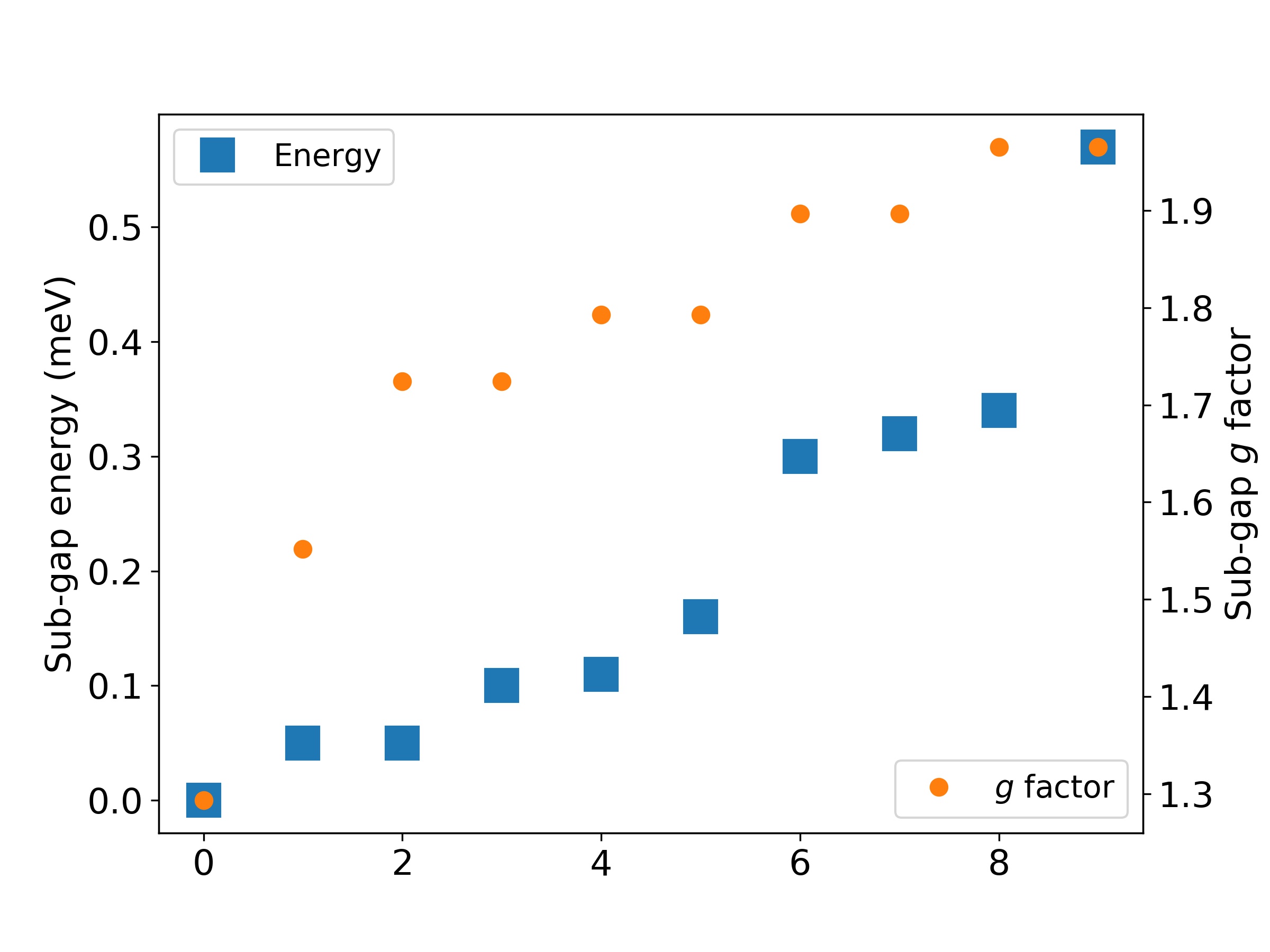}
\caption{\textbf{Multiple dots compilation } Sub-gap zero field energies and $g$ factor for all measured dots.} 
\label{sup: energies and g}
\end{figure}

\begin{figure}
\includegraphics[width = 0.9\textwidth]{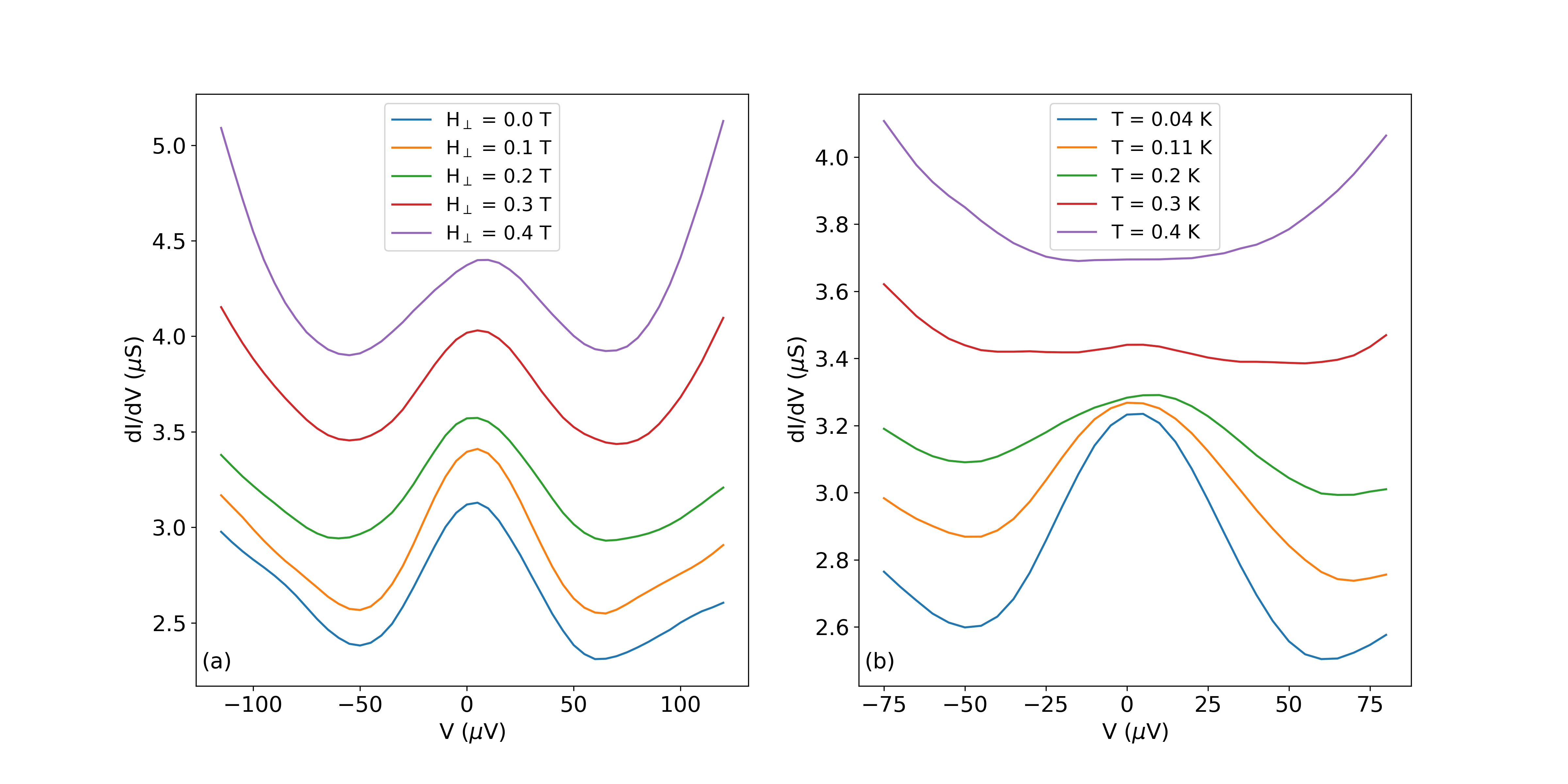}
\caption{\textbf{ZBCP of Device 3:  a.} Perpendicular field dependence of the ZBCP of Device 3. \textbf{b.} Temperature dependence of the ZBCP of Device 3.  } 
\label{sup: ZBCP H and T}
\end{figure}

\end{document}